\documentclass{midl} 
\usepackage{tabularray}
\newcommand{\rev}[1]{\textcolor{black}{#1}}
\newcommand{\reb}[1]{\textcolor{black}{#1}}

\usepackage{mwe} 
\usepackage{graphicx}

\jmlryear{2026}
\jmlrworkshop{Full Paper -- MIDL 2026}
\jmlrvolume{-- 254}
\editors{Accepted for publication at MIDL 2026}

\title[CrossFusion]{CrossFusion: A Multi-Scale Cross-Attention Convolutional Fusion Model for Cancer Survival Prediction}

\midlauthor{
\Name{Rustin Soraki\nametag{$^{1}$}} \Email{rustin@cs.washington.edu}\\
\Name{Huayu Wang\nametag{$^{1}$}} \Email{huayu@uw.edu}\\
\Name{Sitong Liu\nametag{$^{1}$}} \Email{sitonl2@uw.edu}\\
\Name{Joann G. Elmore\nametag{$^{2}$}} \Email{JElmore@mednet.ucla.edu}\\
\Name{Linda Shapiro\nametag{$^{1}$}} \Email{shapiro@cs.washington.edu}\\
\addr $^{1}$ University of Washington, Seattle, WA \\
\addr $^{2}$ University of California, Los Angeles, CA \\
}

\begin{document}

\maketitle

\begin{abstract}
Cancer survival prediction from whole slide images (WSIs) relies on capturing prognostic features spanning multiple magnifications, from global tissue architecture to fine-grained cellular morphology. However, current approaches typically face two main limitations: most frameworks focus heavily on single-scale analysis, thereby overlooking the hierarchical context of tissue; meanwhile, existing multi-scale methods often employ simplistic fusion mechanisms (e.g., direct concatenation) that fail to model effective cross-scale interactions. To address these challenges, we propose CrossFusion, a novel multi-scale architecture that introduces a convolutional fusion processor to perform rigorous scale–space integration. Evaluated on six TCGA cancer cohorts, CrossFusion achieves state-of-the-art C-index performance, consistently outperforming both strong single-scale and multi-scale baselines. Furthermore, leveraging domain-specific pathology feature extractors yields additional gains in prognostic accuracy compared to general-purpose backbones. The source code is available at: \url{https://github.com/RustinS/CrossFusion}
\end{abstract}


\section{Introduction}

Whole Slide Images (WSIs) capture critical tumor characteristics and are central to modern cancer diagnosis~\cite{kumar2020whole,kothari2013pathology,ghaznavi2013digital, liu2025generating}. In clinical practice, survival analysis based on WSIs plays a vital role in informing prognosis and guiding treatment strategies~\cite{campanella2019clinical}. Recent advances in survival analysis have leveraged multiple-instance learning (MIL) and deep learning. For instance, Ilse et al.~\cite{ITW:2018} employ attention mechanisms to identify the most predictive regions, while Li et al.~\cite{li2021dualstreammultipleinstancelearning} aggregate instance-level features into robust slide-level representations. Transformer-based approaches~\cite{shao2021transmil} capture long-range dependencies and graph-based methods~\cite{10.1007/978-3-030-00934-2_20,chen2021whole} model spatial relationships effectively. More recently, methods by Yang et al.~\cite{yang2024scmilsparsecontextawaremultiple} and Wu et al.~\cite{wu2024leveraging} have enhanced interpretability and prediction by integrating sparse attention and prototypical representations. Despite these advancements, the enormous size and heterogeneity of WSIs make it challenging for AI models to capture high-level global tissue patterns and fine-grained cellular detail, both essential for robust survival analysis.

A multi-scale approach offers a promising solution by combining large patches that provide an overview of tissue architecture with small patches that capture detailed cellular morphology. By merging coarse structural cues with fine-grained details, multi-scale methods can better reflect the complex biological processes underlying tumor development and progression. Prior studies have shown that integrating information across multiple scales significantly improves diagnostic accuracy and reduces errors~\cite{tan2023multi}. For example, Deng et al.~\cite{deng2024cross} employ cross-scale attention maps to aggregate features, while Wu et al.~\cite{wu2021scale} use features from different scales as keys, queries, and values to guide learning. Similarly, Zhao et al.~\cite{zhao2024less} select informative patches using a variational positive-unlabeled framework and fuse them with cross-attention. However, these approaches often overlook certain resolution levels or rely on suboptimal fusion techniques, leaving two key challenges unresolved: effectively combining complementary information from multiple scales and developing robust methods for fusing these features.

To address these challenges, we propose \textbf{CrossFusion}, a novel framework that unifies multi-scale patch embeddings from WSIs into a single, predictive representation. We summarize our main contributions as follows:
\begin{enumerate}
    \item \textbf{Multi-Scale Cross-Attention:} This module enables interaction between features at different resolutions, allowing high-resolution details and low-resolution global patterns to reinforce each other while preserving spatial context.
    \item \textbf{Dual-Path Global–Local Context Alignment:} Rather than treating multi-scale fusion as a purely attention-driven or convolution-driven problem, we propose a dual-path global–local alignment mechanism, where transformers capture cross-scale global dependencies convolutions enforce spatial alignment and local coherence.
    \item \textbf{Extensive Validation \& Accuracy:} We validate CrossFusion on diverse cancer survival datasets, demonstrating that it matches or exceeds state-of-the-art performance in survival analysis while maintaining interpretability through visualization of key regions at multiple magnifications. We also examine the effect of different feature extraction backbones and compare CrossFusion trained on the domain-specific backbones to the general one.
\end{enumerate}

\section{Related Work}
To process these gigapixel-resolution images, Multiple Instance Learning (MIL) has become the standard paradigm. In this framework, a WSI is treated as a "bag" of smaller patches (instances), and a slide-level prediction is aggregated from patch-level features. While early methods focused on simple aggregation, recent advancements have introduced attention mechanisms, graph convolutional networks (GCNs), and transformers to better capture the spatial and semantic dependencies between tissue patches.
\subsection{Single-Scale Computational Pathology Methods}
Most state-of-the-art methods operate on a single magnification scale. Attention-based models like AMIL~\cite{ITW:2018} and DSMIL~\cite{li2021dualstreammultipleinstancelearning} aggregate patch features by identifying predictive regions. Graph-based approaches, such as DeepGraphSurv~\cite{10.1007/978-3-030-00934-2_20} and Patch-GCN~\cite{chen2021whole}, model spatial topology , while Transformer-based methods like TransMIL~\cite{shao2021transmil} capture long-range dependencies. We also compare against recent specialized architectures like SCMIL~\cite{yang2024scmilsparsecontextawaremultiple} and ProtoSurv~\cite{wu2024leveraging}, which utilize sparse attention and prototypical representations to enhance interpretability and performance.

\subsection{Multi-Scale Histopathology Methods}

Multi-scale methods aim to mimic the pathologist's workflow by integrating coarse structural cues with fine-grained cellular details. \rev{Existing frameworks have adopted distinct strategies to achieve this: ZoomMIL~\cite{thandiackal2022differentiable} selectively mines salient regions for high-magnification feature extraction, whereas CSMIL~\cite{deng2024cross} employs convolutional networks to fuse multi-scale features. Additionally, MuSTMIL~\cite{marini2021multi} leverages multi-task learning to maximize data utilization, and \rev{HIPT~\cite{HIPT}} utilizes a feature pyramid approach to derive hierarchical representations for survival tasks.}

\rev{However, while prior works have employed mechanisms such as cross-scale attention maps or scale-specific key-query interactions, they often overlook explicit connections between resolutions or rely on suboptimal fusion techniques. CrossFusion addresses these limitations by using multi-magnification patch features to explicitly model ``intra-scale and inter-scale'' interactions. Through a novel integration of cross-scale cross-attention, convolutional fusion, and Transformer encoding, our framework directly outputs time-series survival risks, demonstrating significant performance gains over existing methods across multiple cancer cohorts.}

\section{Method}

\begin{figure}[t]
    \centering
    \includegraphics[width=1\textwidth]{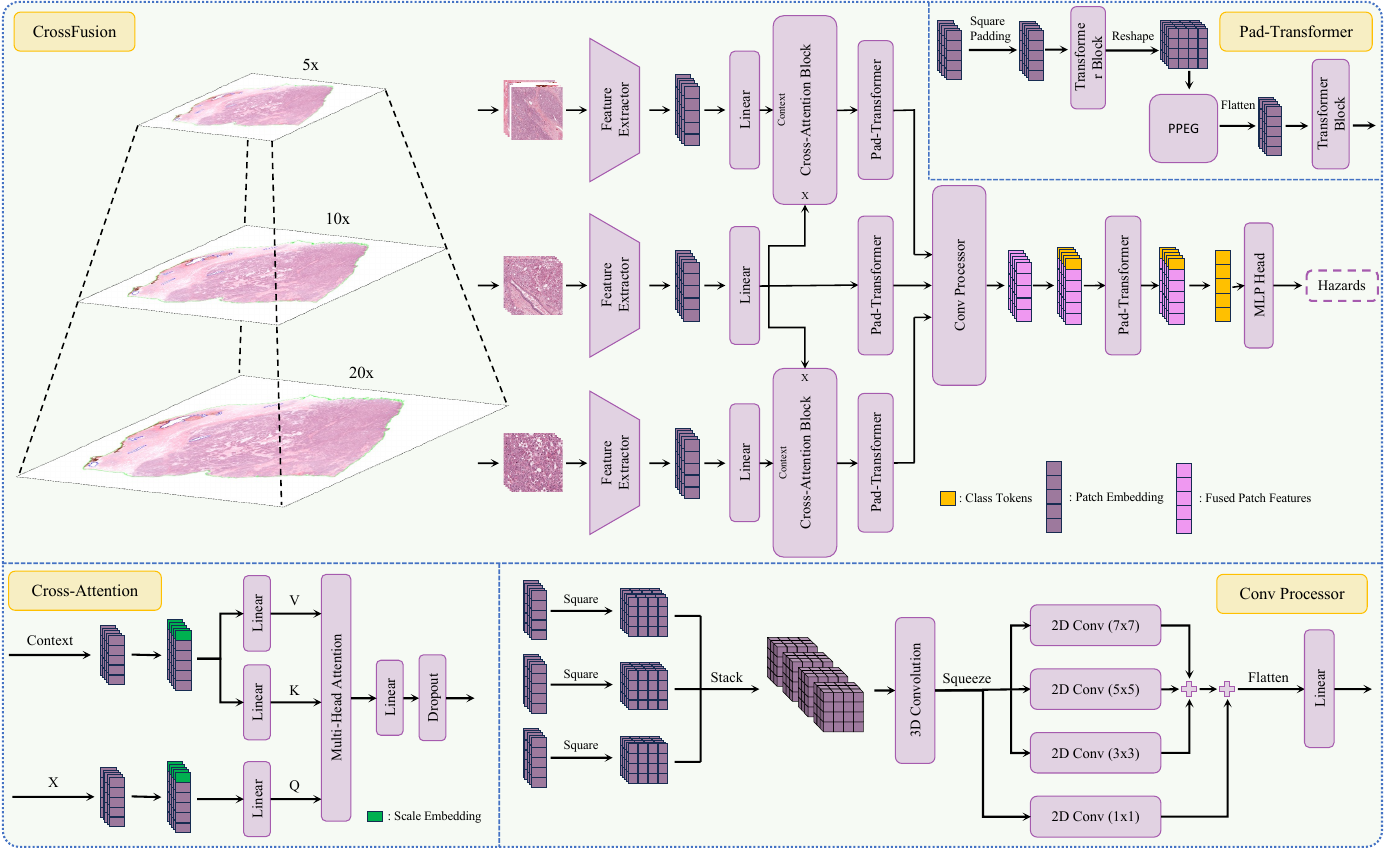}
    \caption{Overview of CrossFusion. WSIs are processed by extracting patches at 5x (coarse), 10x (source), and 20x (fine) magnifications, which are first encoded using a feature extractor and then projected into a common embedding space. The source features interact with the coarse and fine features via cross-attention blocks, and each branch is refined by Pad-Transformers. The multi-scale features are subsequently fused using a Conv Processor, and a replicated learnable class token is appended. An additional transformer block refines this token, and an MLP head produces the final survival predictions from the class tokens.}
    \label{fig:crossfusion}
\end{figure}

This section outlines the pipeline of our proposed methodology, CrossFusion. Figure \ref{fig:crossfusion} illustrates the complete framework, including its main stages and components. As shown in the figure, the extracted patches at different magnifications are encoded by a feature extractor, projected into a common embedding space, and then through the Cross-Attention Block, in which the different magnifications interact. The next step is the Pad-Transformer process, which uses Pyramid Position Encoding to capture local and global context. The Conv Processor is used to fuse the multi-scale features.  

\subsection{CrossFusion}

The CrossFusion module takes three inputs: \(\mathbf{X}_{\text{C}}\) (coarse), \(\mathbf{X}_{\text{S}}\) (source), and \(\mathbf{X}_{\text{F}}\) (fine), which represent patch embeddings from 5x, 10x, and 20x magnifications, respectively. Initially, each embedding is projected into a shared space \(D_e\). Next, to facilitate inter-scale interactions, the module applies cross-attention with \(\mathbf{X}_{\text{S}}\) as the query and the other embeddings as context:

\begin{equation}
    \mathbf{X'}_{\text{C}} = \text{\textit{CAB}}(\mathbf{X}_{\text{S}}, \mathbf{X}_{\text{C}}), \quad
    \mathbf{X'}_{\text{F}} = \text{\textit{CAB}}(\mathbf{X}_{\text{S}}, \mathbf{X}_{\text{F}})
\end{equation}

\noindent where \textit{CAB} denotes the Cross-Attention Block. Each feature set is then processed by dedicated Pad-Transformer (\textit{PT}) blocks. The outputs are fused via the Conv Processor (\textit{CP}):

\begin{equation}
    \mathbf{X}_{\text{fused}} = \text{\textit{CP}}\Big(\text{\textit{PT}}(\mathbf{X'}_{\text{C}}),\, \text{\textit{PT}}(\mathbf{X}_{\text{S}}),\, \text{\textit{PT}}(\mathbf{X'}_{\text{F}})\Big)
\end{equation}

A learnable class token \(\mathbf{c} \in \mathbb{R}^{1 \times D_e}\) is first replicated and prepended to the fused token sequence. This extended sequence is processed by an additional Pad-Transformer followed by Layer Normalization. The class token is then extracted to yield \(\mathbf{c'}\). An MLP head maps \(\mathbf{c'}\) to logits \(\mathbf{l}\), from which hazards \(\mathbf{h}\) are computed using a sigmoid activation. Finally, survival probabilities \(\mathbf{S}\) are obtained as the cumulative product of \(1 - \mathbf{h}\).

\subsection{Cross-Attention Block}  

The cross-attention module fuses information from two inputs: the primary input  
\(\mathbf{X} \in \mathbb{R}^{B \times N \times D}\) and a contextual input  
\(\mathbf{Context} \in \mathbb{R}^{B \times M \times D}\), where \(B\) is the batch size, \(N\) and \(M\) are sequence lengths, and \(D\) is the embedding dimension. Both inputs are augmented with a learnable scale embedding  
\(\mathbf{s} \in \mathbb{R}^{1 \times 1 \times D}\) before computing attention, acting like a learnable positional encoding. The queries, keys, and values are obtained via linear projections:  

\begin{equation}
  \mathbf{Q} = W_q (\mathbf{X} + \mathbf{s}), \quad
  \mathbf{K} = W_k (\mathbf{Context} + \mathbf{s}), \quad
  \mathbf{V} = W_v (\mathbf{Context} + \mathbf{s}),
\end{equation}

\noindent where \(W_q, W_k, W_v \in \mathbb{R}^{D \times D}\) are learnable weight matrices. Multi-head attention is then applied, followed by an output projection. The output is further processed through residual connections, layer normalization, and a feed-forward network. This mechanism enables effective information exchange between the primary input and its contextual counterpart, improving feature representation.  

\subsection{Pad-Transformer}

The Pad-Transformer organizes input tokens into a grid, processes them with an initial Transformer block, uses Pyramid Position Encoding Generator (PPEG)~\cite{shao2021transmil} to add spatial context, and then further refines the features using a second Transformer block. This combines global dependencies (from the first block) and local details (via PPEG) before final refinement by the second Transformer block.

\subsubsection{Square Padding}  
Given an input sequence \(\mathbf{X} \in \mathbb{R}^{B \times N \times D}\), where \(N\) is the number of tokens, we first compute \(H = W = \lceil \sqrt{N} \rceil\). We then pad \(\mathbf{X}\) by appending the first \((H \times W) - N\) tokens to the end of the sequence, resulting in a sequence of length \(H \times W\). Finally, this padded sequence is reshaped into a square grid \(\mathbf{X}_s \in \mathbb{R}^{B \times H \times W \times D}\) for subsequent spatial operations.

\subsubsection{Transformer Blocks}  
Each transformer block applies layer normalization, computes multi-head self-attention, and adds the result to the input via a residual connection. A second layer normalization is followed by a feed-forward network, GELU activation, and dropout. 

\subsubsection{Convolutional Positional Encoding (PPEG)}  
To incorporate local spatial context, the PPEG module applies three parallel depth-wise convolutions with kernel sizes \(7\), \(5\), and \(3\) to the reshaped feature map  
\(\mathbf{X}_s \in \mathbb{R}^{B \times D \times H \times W}\):

\begin{equation}
\mathbf{X}_{\text{PPEG}} = \text{Conv}_7(\mathbf{X}_s) + \text{Conv}_5(\mathbf{X}_s) + \text{Conv}_3(\mathbf{X}_s) + \mathbf{X}_s
\end{equation}

This operation enhances token representations with detailed local positional information.

\subsection{Conv Processor}  
The Conv Processor fuses multi-source features and enhances spatial representations using multi-scale convolutions. The combination of Transformer and CNN leverages global and local modeling strengths: Transformer captures long-range dependencies while CNN reinforces local spatial features afterward, enabling multi-source information fusion and compensating for each other's limitations.  Given three input sequences \(\mathbf{X}_i \in \mathbb{R}^{B \times N \times D}, \quad i\in\{1,2,3\}\), each is square-padded and reshaped into a 2D feature map, \(\mathbf{X}_i' \in \mathbb{R}^{B \times D \times H \times W},\) where \(N = H \times W\). The feature maps are stacked into \(\mathbf{X}_{\text{stack}} \in \mathbb{R}^{B \times 3 \times D \times H \times W}\) and fused via a 3D convolution to get \(\mathbf{X}_{\text{fused}}\). After squeezing the singleton channel, multi-scale features are extracted by applying parallel depth-wise convolutions with kernel sizes \(7\), \(5\), \(3\), and \(1\), where the output dimension is reduced to \(D' = D // 2\):

\begin{equation}
\mathbf{X}_{\text{ms}} = \text{Conv}_7(\mathbf{X}_{\text{fused}}) + \text{Conv}_5(\mathbf{X}_{\text{fused}}) + \text{Conv}_3(\mathbf{X}_{\text{fused}}) + \text{Conv}_1(\mathbf{X}_{\text{fused}}).
\end{equation}

The resulting feature map is flattened along spatial dimensions into \(\mathbf{X}_{\text{flat}} \in \mathbb{R}^{B \times D' \times (H \cdot W)}\) and permuted into a token sequence \(\mathbf{X}_{\text{seq}} \in \mathbb{R}^{B \times (H \cdot W) \times D'}\). Finally, a linear projection followed by Layer Normalization restores the original dimension ($D$). This module efficiently fuses multi-source information while capturing multi-scale spatial features.

\section{Experimental Setup}

\subsection{Dataset}
We used H\&E WSIs from six TCGA cancer types: BLCA (437 slides), BRCA (1016 slides), COAD (424 slides), GB\&LGG (1041 slides), LUAD (507 slides), and UCEC (539 slides). These datasets were chosen for their size, public availability, survival follow-up data, and a balanced uncensored-to-censored ratio (average 0.28). On average, each WSI yields 13,496 patches at 20x, 3,449 patches at 10x, and 895 patches at 5x, with the number of 20x patches reaching up to 137,990.

\subsection{Implementation Details}
\paragraph{Patch Extraction and Embedding:}  
We used CLAM~\cite{lu2021data} to extract 256×256 patches at 20x, 10x, and 5x magnifications and extract features from different feature extraction backbones. Tissue regions were identified using a binary mask computed by thresholding the saturation channel in HSV.

\paragraph{Training and Evaluation:}  
The model was trained with Adam (learning rate \(1\times10^{-4}\), weight decay \(4\times10^{-6}\), batch size 1) with a 5-epoch warm-up, and evaluated via 5-fold cross-validation. For a fair comparison, all methods used the same loss function, feature embeddings, and hyperparameters. Experiments were implemented in PyTorch on a workstation with four Nvidia RTX A4000 GPUs.

\paragraph{Evaluation Metrics:}  
Performance was measured using the mean C-index across validation splits. Additionally, we report the p-value from stratifying patients into high- and low-risk groups as a statistical measure of the model’s discriminative ability.

\section{Experiments and Results}
In this section, we evaluate our model's performance through experiments. First, we compare CrossFusion to state-of-the-art methods. Next, we assess our model's interpretability by analyzing attention-based heatmaps, providing insights into its decision-making process. Finally, we analyze the effect of using different foundational models as feature extraction backbones to determine whether domain-specialized backbones improve performance.

\subsection{Comparison with State-Of-The-Art Methods}

\begin{table*}[t]
    \centering
    \fontsize{8}{10}\selectfont
    \caption{C-Index ($\text{mean}_{\ \text{std}}$) of different methods over the six different datasets. The best and the second-best results are highlighted in \textbf{bold} and \underline{underline}, respectively.}
    \begin{tblr}{
      colspec={lcccccc},
      hline{1} = {1pt,solid},
      rowsep=1.2pt,
      colsep = 2.5pt,
    }
    { } & BLCA & BRCA & COAD & GB\&LGG & LUAD & UCEC \\ 
    \hline
    \SetCell[c=7]{l,gray9} \textit{Single Scale Methods} & & & & & & \\
    AMIL & $.559_{.059}$ & $.590_{.050}$ & $.662_{.063}$ & $.759_{.111}$ & $.590_{.036}$ & $.644_{.092}$ \\ 
    DSMIL & $.552_{.050}$ & $.564_{.044}$ & $.610_{.012}$ & $.728_{.102}$ & $.579_{.032}$ & $.601_{.073}$ \\ 
    TransMIL & $.574_{.064}$ & $.594_{.045}$ & $.656_{.057}$ & $.772_{.093}$ & $.594_{.059}$ & $.664_{.044}$ \\ 
    DeepGraphSurv & $.572_{.054}$ & $.558_{.099}$ & $.591_{.119}$ & $.764_{.053}$ & $.622_{.055}$ & $.635_{.061}$ \\
    PatchGCN & $.563_{.043}$ & $.595_{.089}$ & $.612_{.144}$ & $.774_{.046}$ & $.577_{.081}$ & $.679_{.071}$ \\
    SCMIL & $.566_{.054}$ & $.590_{.034}$ & $\underline{.677_{.070}}$ & $.763_{.094}$ & $.584_{.050}$ & $.668_{.071}$ \\ 
    ProtoSurv & $.579_{.023}$ & $.627_{.034}$ & $.668_{.057}$ & $.776_{.031}$ & $.619_{.046}$ & $\mathbf{.730_{.032}}$ \\
    \hline[dotted]
    \SetCell[c=7]{l,gray9} \textit{Multi Scale Methods} & & & & & & \\
    ZoomMIL & $.570_{.056}$ & $.563_{.047}$ & $.642_{.066}$ & $.770_{.091}$ & $.568_{.046}$ & $.679_{.033}$ \\
    MuSTMIL & $.575_{.065}$ & $.589_{.043}$ & $.640_{.084}$ & $.780_{.089}$ & $.600_{.040}$ & $.682_{.039}$ \\
    CSMIL & $.542_{.071}$ & $.589_{.070}$ & $.636_{.087}$ & $.742_{.119}$ & $.582_{.060}$ & $.640_{.047}$ \\
    
    \rev{HIPT} & \rev{$.554_{.015}$} & \rev{$.603_{.045}$} & \rev{$.651_{.102}$} & \rev{$.773_{.032}$} & \rev{$.572_{.035}$} & \rev{$.697_{.060}$} \\
    
    \hline[dotted]
    \SetCell[c=7]{l,gray9} \textit{Ours} & & & & & & \\
    CrossFusion w/o CP & $\underline{.627_{.014}}$ & $\underline{.631_{.076}}$ & $.669_{.069}$ & $\underline{.787_{.081}}$ & $\underline{.627_{.038}}$ & $\underline{.710_{.061}}$ \\ 
    CrossFusion w/o F\&C & $.562_{.058}$ & $.629_{.052}$ & $.631_{.052}$ & $.782_{.074}$ & $.609_{.053}$ & $.672_{.050}$ \\ 
    \textbf{CrossFusion} & $\mathbf{.630_{.027}}$ & $\mathbf{.643_{.037}}$ & $\mathbf{.694_{.053}}$ & $\mathbf{.797_{.056}}$ & $\mathbf{.627_{.040}}$ & $.702_{.044}$ \\ 
    \hline
    \end{tblr}
    \label{tab:sota-compare}
\end{table*}
We evaluated CrossFusion against state-of-the-art survival prediction methods, categorized into single-scale and multi-scale approaches. For single-scale methods, we compared against AMIL~\cite{ITW:2018}, DSMIL~\cite{li2021dualstreammultipleinstancelearning}, TransMIL~\cite{shao2021transmil}, DeepGraphSurv~\cite{10.1007/978-3-030-00934-2_20}, Patch-GCN~\cite{chen2021whole}, SCMIL~\cite{yang2024scmilsparsecontextawaremultiple}, and ProtoSurv~\cite{wu2024leveraging}. For multi-scale methods, which aim to mimic the pathologist's workflow by integrating coarse structural cues with fine-grained cellular details, we compared against ZoomMIL~\cite{thandiackal2022differentiable}, MUSTMIL~\cite{marini2021multi}, \rev{HIPT~\cite{HIPT}}and CSMIL~\cite{deng2024cross}. All models used ResNet50~\cite{He_2016_CVPR} as the feature extractor for fair comparison.

As shown in Table \ref{tab:sota-compare}, CrossFusion achieves the best or near-optimal performance across six cancer datasets. In the UCEC dataset, the low uncensored-to-all-slides ratio (0.15) posed a challenge due to CrossFusion’s reliance on patch-level features without prior information, resulting in slightly lower performance than ProtoSurv, which leverages priors. Nevertheless, CrossFusion consistently outperforms all other baselines and remains competitive with ProtoSurv, demonstrating robustness even under data constraints. Comparing with other multi-scale model, the superior performance suggests that CrossFusion functions not merely as a feature aggregator but as a scale-interrogative learner, enforcing consistency checks between tissue organization and cellular morphology under the intermediate 10x view.

Ablation studies validated the contributions of key components. First, replacing the ConvProcessor (CP) with simple concatenation/projection reduced performance on all datasets except UCEC (due to data scarcity), validating the effectiveness of our Global–Local Context Alignment module. Second, removing Fine (F) and Coarse (C) sources to rely solely on 20× patches—following mainstream single-scale approaches like MMP~\cite{mmp} and UniPro~\cite{unipro}—significantly degraded performance in most datasets, confirming the necessity of multi-scale inputs for capturing discriminative WSI features. 

Finally, stratification analysis yielded p-values of \(1.79 \times 10^{-4}\) for BLCA, \(1.49 \times 10^{-2}\) for BRCA, \(3.30 \times 10^{-4}\) for COAD, \(2.30 \times 10^{-39}\) for GB\&LGG, \(1.92 \times 10^{-2}\) for LUAD, and \(3.91 \times 10^{-3}\) for UCEC. These statistically significant results confirm that CrossFusion effectively differentiates high-risk and low-risk patient groups, underscoring its clinical relevance for survival prediction.

\begin{figure*}[t]
\centering
\begin{minipage}[t]{0.32\textwidth}
  \centering
  \includegraphics[width=\linewidth]{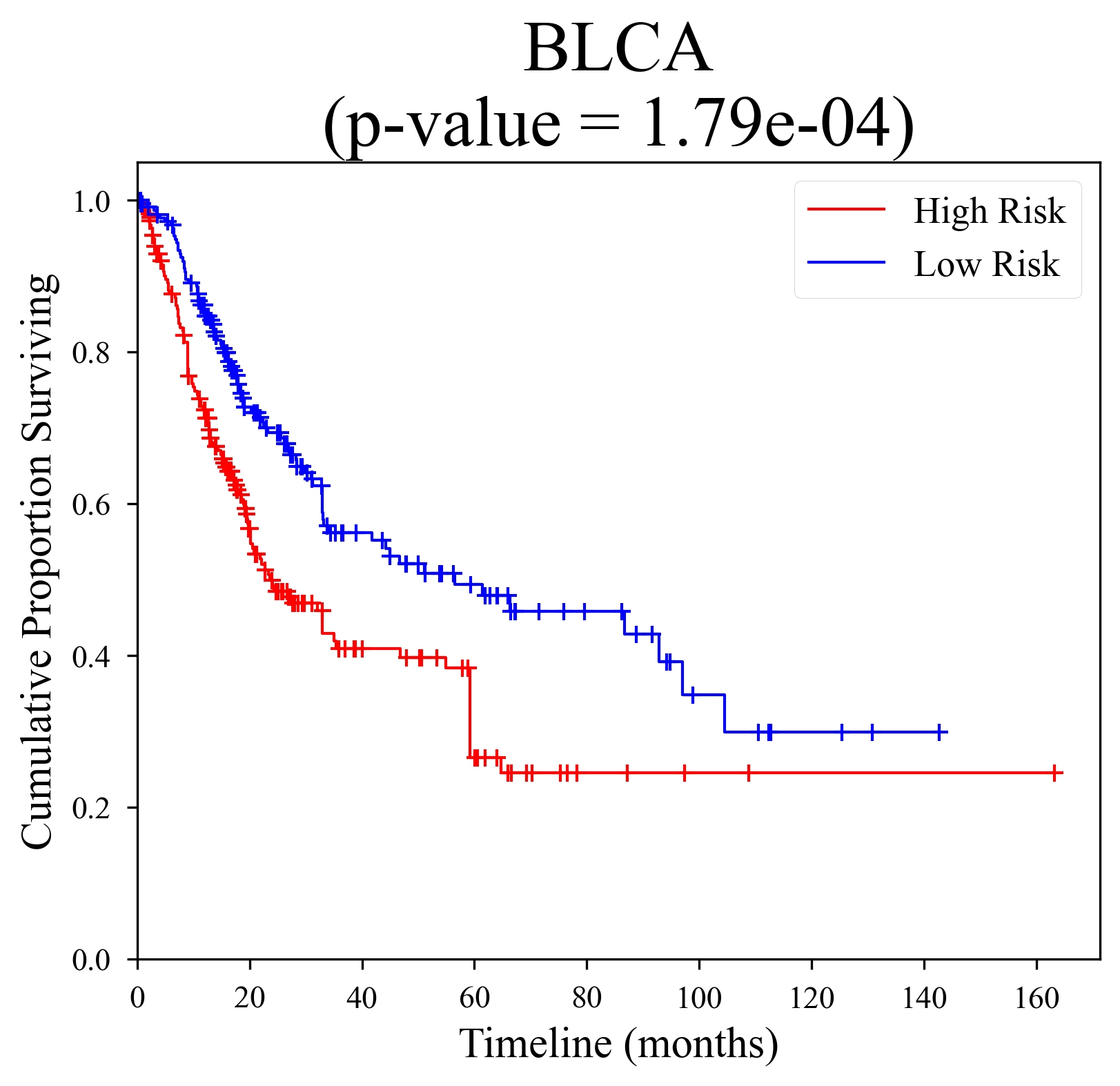}
  \label{fig:km_blca}
\end{minipage}\hfill
\begin{minipage}[t]{0.32\textwidth}
  \centering
  \includegraphics[width=\linewidth]{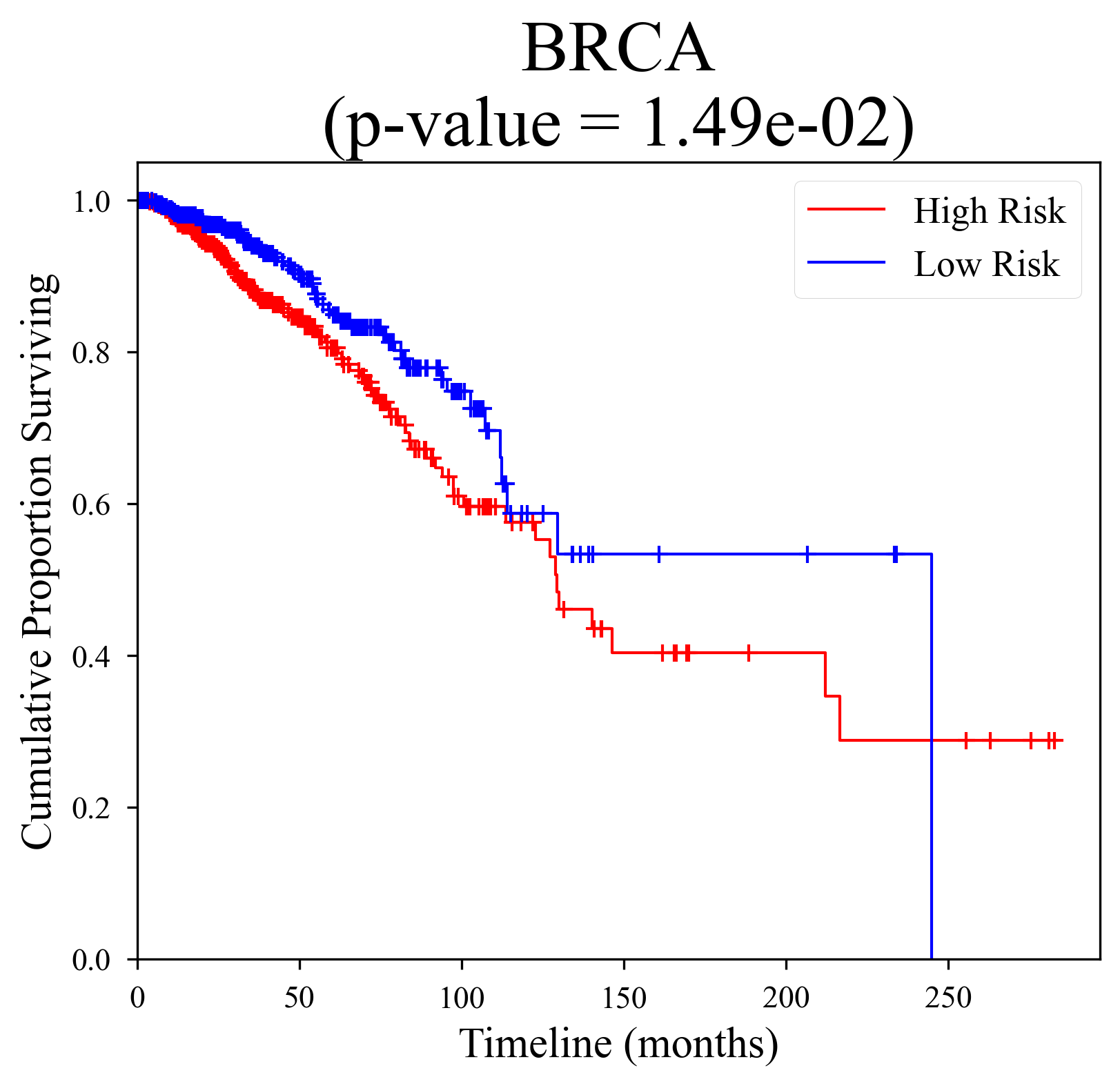}
  \label{fig:km_brca}
\end{minipage}\hfill
\begin{minipage}[t]{0.32\textwidth}
  \centering
  \includegraphics[width=\linewidth]{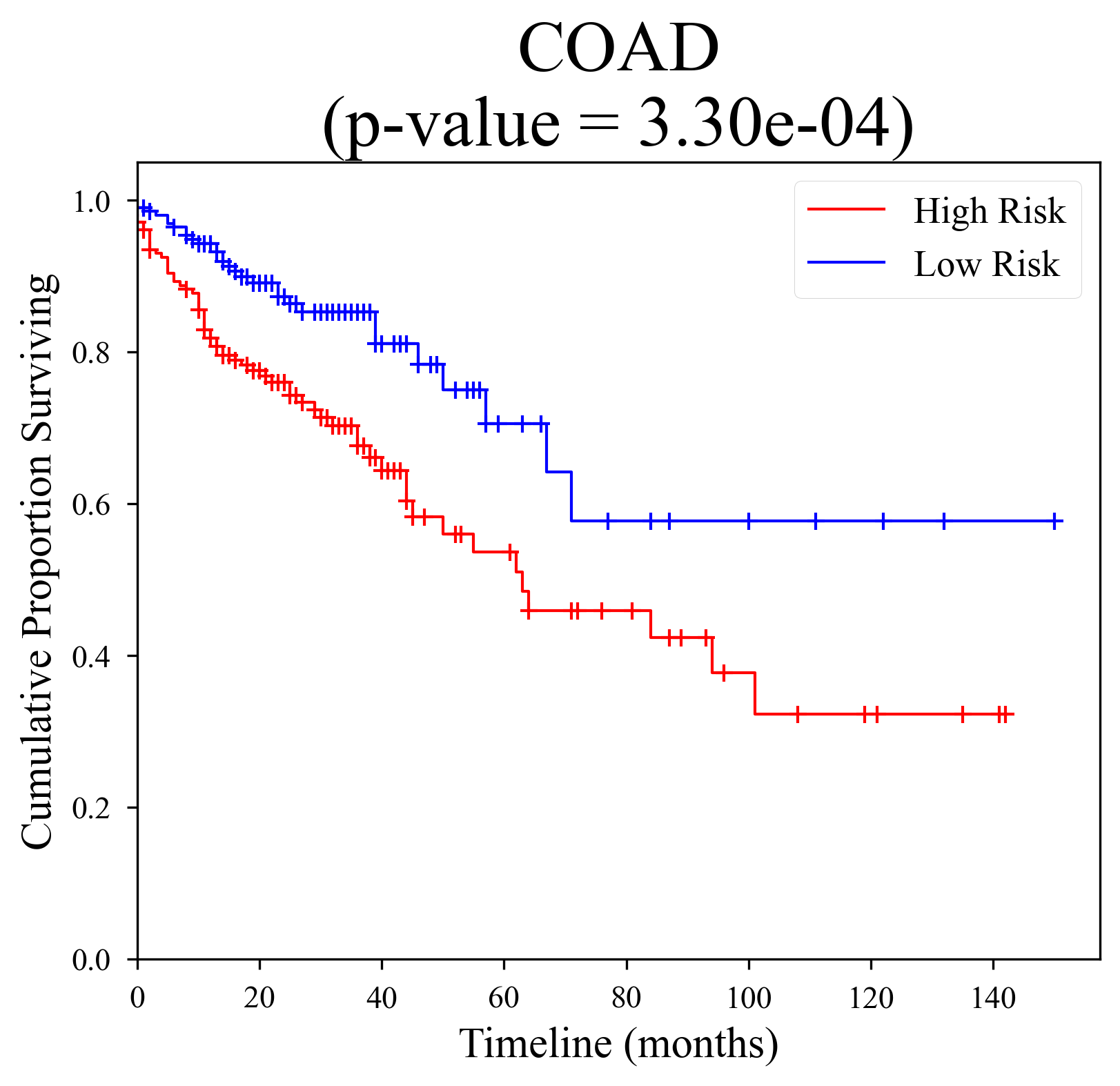}
  \label{fig:km_coad}
\end{minipage}

\vspace{0.6em}

\begin{minipage}[t]{0.32\textwidth}
  \centering
  \includegraphics[width=\linewidth]{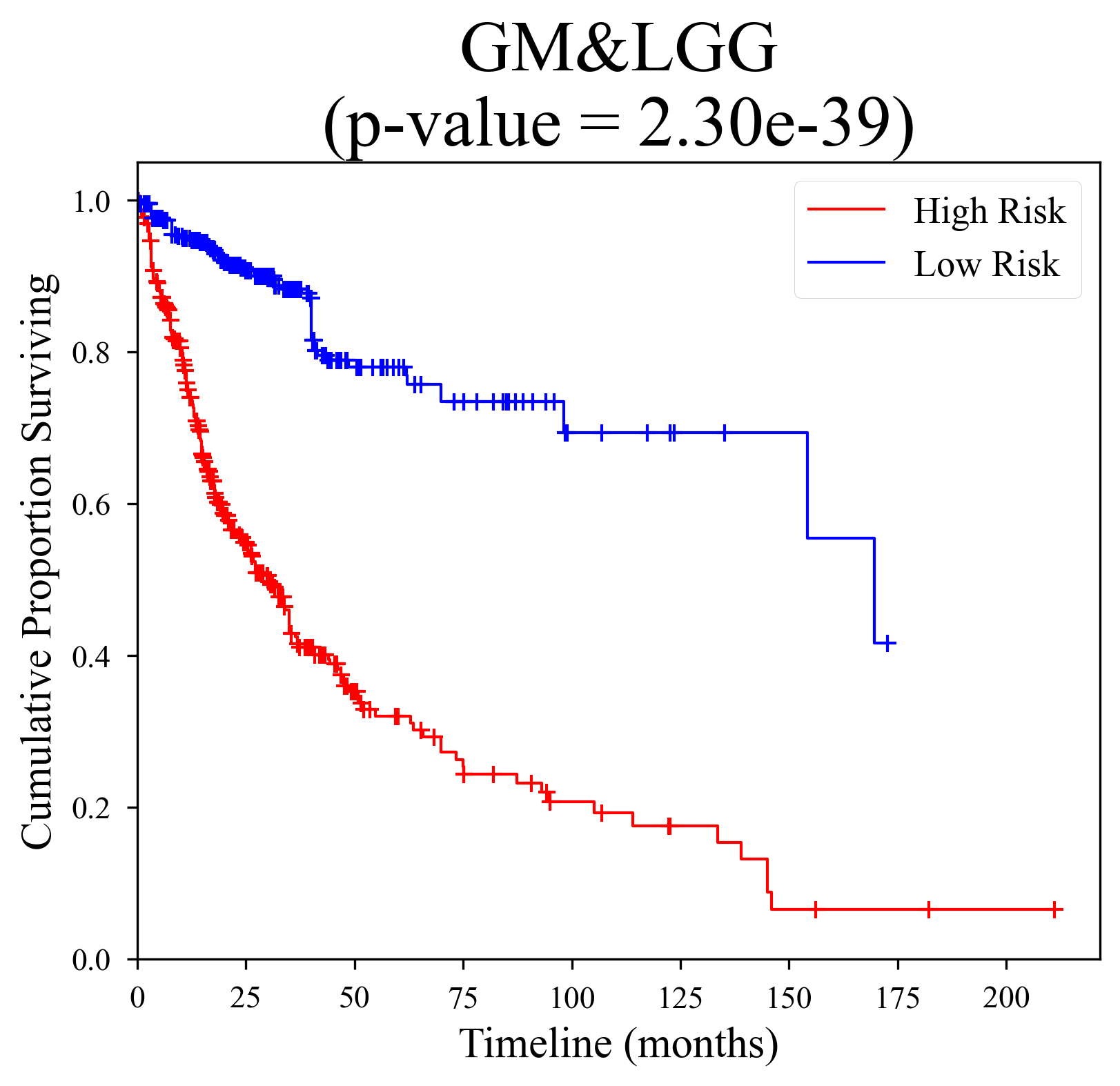}
  \label{fig:km_gmlgg}
\end{minipage}\hfill
\begin{minipage}[t]{0.32\textwidth}
  \centering
  \includegraphics[width=\linewidth]{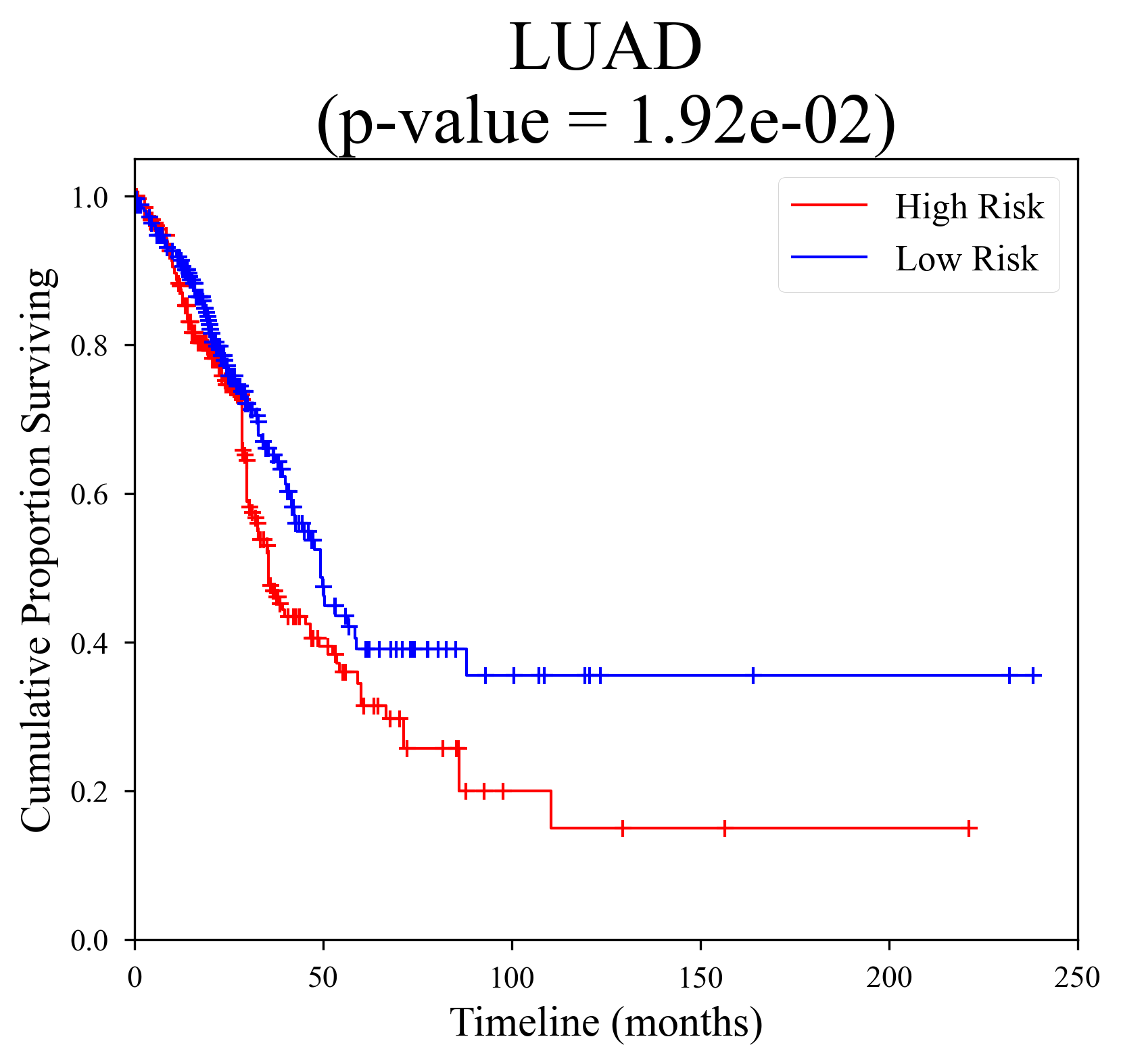}
  \label{fig:km_luad}
\end{minipage}\hfill
\begin{minipage}[t]{0.32\textwidth}
  \centering
  \includegraphics[width=\linewidth]{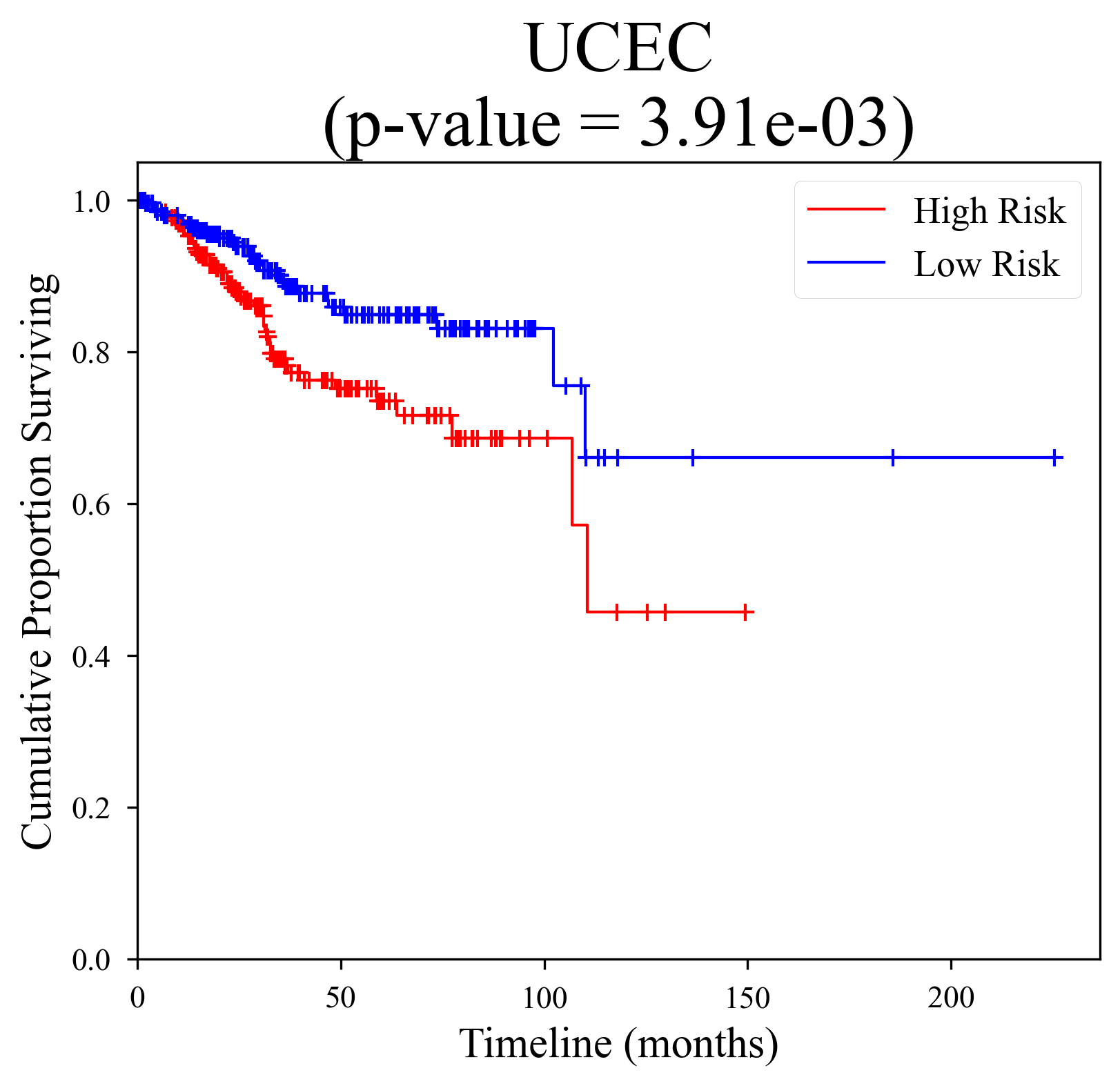}
  \label{fig:km_ucec}
\end{minipage}

\caption{\reb{\textbf{Kaplan--Meier survival curves.} Kaplan--Meier curves of predicted high-risk (red) and low-risk (blue) groups across six TCGA cohorts. The log-rank test p-value is reported in each panel; $p<0.05$ indicates statistically significant separation.}}
\label{fig:km_all}
\end{figure*}

\reb{We further assessed CrossFusion’s prognostic stratification by splitting patients into predicted high and low risk groups and visualizing their survival trajectories using Kaplan–Meier curves. As shown in Figure \ref{fig:km_all}, the low-risk group (blue) consistently exhibits higher survival probabilities over time than the high-risk group (red) across all six TCGA cohorts (BLCA, BRCA, COAD, GM\&LGG, LUAD, UCEC). The separation between the two curves is statistically significant in each cohort based on the log-rank test (all $p$-values $<$ 0.05), with particularly strong discrimination observed in GM\&LGG and additional clear separation in BLCA and COAD. These results indicate that CrossFusion learns risk scores that meaningfully stratify patients into subgroups with distinct survival outcomes.}

\subsection{Interpretability}  
\begin{figure}[t]
    \centering
    \includegraphics[width=\textwidth]{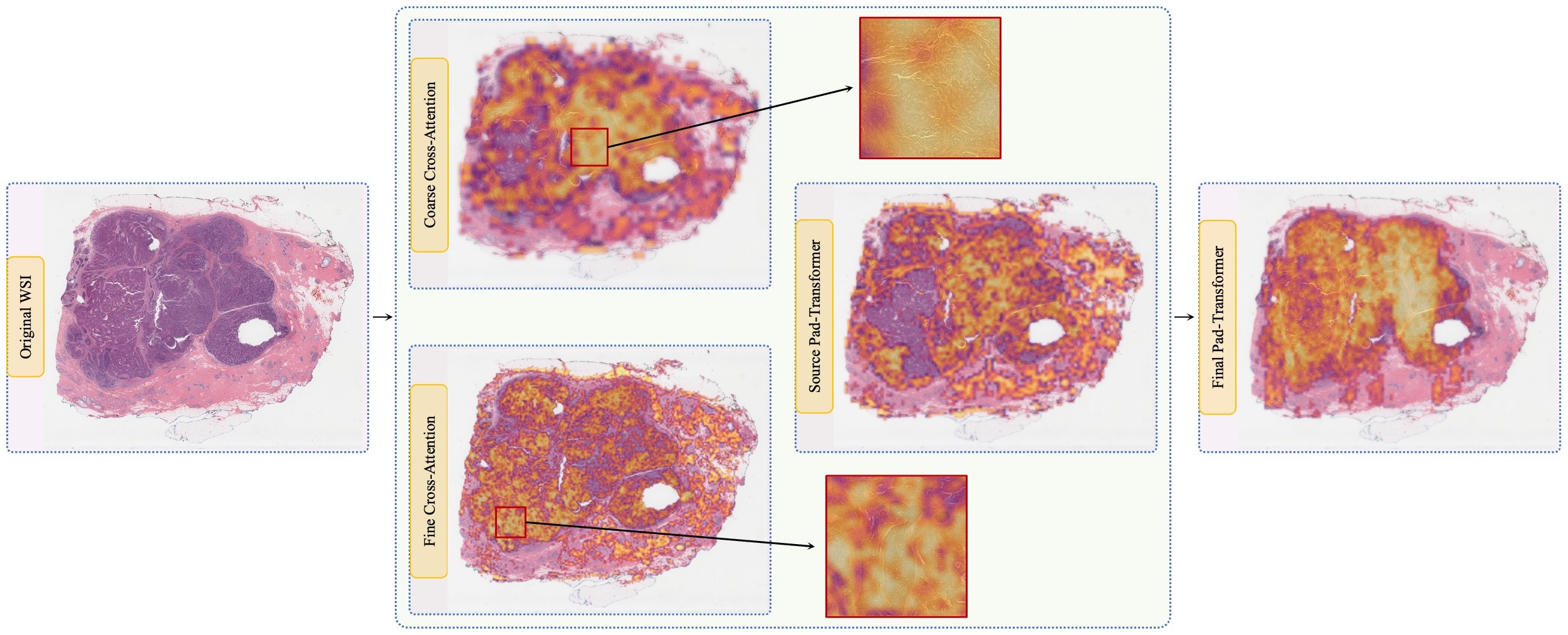}
    \caption{Generated heatmaps from the model predicting a high-risk case. The dark purple clusters mark tumor regions in the original WSI on the left and lighter yellow areas highlight important regions from the model's attention weights.} 
    \label{fig:heatmaps}
\end{figure}

To explore the model's decision-making, we generated heatmaps from attention weights in the Cross-Attention layers, the source features Pad-Transformer, and the fused features Pad-Transformer. Figure \ref{fig:heatmaps} shows a WSI from the TCGA-BRCA dataset—depicting a high-risk patient with low survival time—alongside its corresponding heatmaps.

The three intermediate heatmaps reveal that different modules capture distinct features: the Coarse Cross-Attention layer focuses on large-scale tissue organization, the Fine Cross-Attention layer captures detailed cellular morphology, and the Source Pad-Transformer emphasizes intermediate-scale structures. The final heatmap from the last transformer layer demonstrates that the model effectively filters out less relevant regions, concentrating on key histopathological features.  

\subsection{Analyzing the effect of Different Feature Extraction Backbones}

We evaluate CrossFusion using different feature extraction backbones, comparing their impact on model performance. Specifically, we extract patch-level features using Conch~\cite{lu2024avisionlanguage}, Uni2-h~\cite{chen2024uni}, QuiltNet~\cite{ikezogwo2023quilt1m}, and Prov-GigaPath~\cite{xu2024gigapath}, and compare them against features extracted using ResNet50.  

Table \ref{tab:extract-compare} shows that CrossFusion performs best with features from the Uni2-h backbone, while other domain-specific backbones yield similar results. The performance gap is highlighted particularly in the BRCA and the UCEC datasets, where utilizing high-quality features is crucial because of the low uncensored-to-
all-slides ratio. The clear performance gap between CrossFusion trained on specialized backbones and CrossFusion trained on ResNet50 backbone highlights the benefit of domain-specific extraction backbones, which better capture tissue-level details, such as cellular morphology and tissue architecture, crucial for accurate prognostication.

\rev{While the transition to the Uni2-h backbone provides a general performance uplift across all methods, CrossFusion demonstrates the distinct ability to maximize the potential of these advanced representations. As shown in Table \ref{tab:uni2}, our method consistently outperforms competing approaches under the same experimental settings. Specifically, CrossFusion not only surpasses the ResNet-based HIPT baseline by a substantial margin of 5.7\% but, more importantly, maintains a clear lead over other multi-scale methods equipped with the same Uni2-h backbone (e.g., surpassing MuSTMIL by an average of 2.6\%). This consistent superiority—particularly the 8.1\% and 7.4\% gains on BRCA and BLCA against HIPT—confirms that the performance leap stems from our effective cross-scale fusion mechanism, rather than solely from the upgrade in feature extractors.}

\begin{table*}[h]
    \centering
    \fontsize{8}{10}\selectfont
    \caption{C-Index ($\text{mean}_{\ \text{std}}$) of \textbf{CrossFusion} trained on different feature extraction backbones over the six different datasets. The best and the second-best results are highlighted in \textbf{bold} and \underline{underline}, respectively.}
    \begin{tblr}{
      colspec={lcccccc|c}, 
      hline{1} = {1pt,solid},
      rowsep=1.2pt, 
      colsep = 2.5pt,
    }
    { } & BLCA & BRCA & COAD & GB\&LGG & LUAD & UCEC & Mean \\ 
    \hline
    w/ ResNet50 {\tiny(Base)} & $.630_{.027}$ & $.643_{.037}$ & $.694_{.053}$ & $.797_{.056}$ & $\underline{.627_{.040}}$ & $.702_{.044}$ & $.682$ \\ 
    \hline[dotted]
    w/ Conch & $\mathbf{.649_{.053}}$ & $.675_{.060}$ & $.705_{.024}$ & $.799_{.055}$ & $.604_{.055}$ & $\underline{.737_{.043}}$ & $.695$ \\ 
    
    w/ Uni2-h & $.628_{.019}$ & $\underline{.684_{.043}}$ & $.698_{.023}$ & $\underline{.810_{.067}}$ & $.625_{.051}$ & $\mathbf{.745_{.030}}$ & $\mathbf{.698}$  \\ 
    
    w/ QuiltNet & $.614_{.050}$ & $.650_{.017}$ & $\underline{.712_{.043}}$ & $\mathbf{.812_{.032}}$ & $
    \mathbf{.640_{.064}}$ & $.727_{.028}$ & $.693$ \\ 
    
    w/ Prov-GigaPath & $\underline{.635_{.023}}$ & $\mathbf{.686_{.037}}$ & $\mathbf{.718_{.064}}$ & $.802_{.051}$ & $.620_{.063}$ & $.724_{.043}$ & $\underline{.698}$ \\ 
    \hline
    \end{tblr}
    \label{tab:extract-compare}
\end{table*}

\begin{table*}[t]
    \centering
    \fontsize{8}{10}\selectfont
    \caption{\rev{Performance comparison (C-Index $\text{mean}_{\ \text{std}}$) of \textbf{CrossFusion} against other multi-scale MIL methods using \textbf{Uni2-h} features across six datasets.}}
    \begin{tblr}{
      colspec={lcccccc}, 
      hline{1} = {1pt,solid},
      hline{2} = {0.5pt,solid}, 
      hline{5} = {1pt,solid},   
      rowsep=3pt, 
      colsep=4pt,
    }
    { } & BLCA & BRCA & COAD & GB\&LGG & LUAD & UCEC \\ 
    
    MuSTMIL     & $.624_{.047}$ & $.634_{.037}$ & $.682_{.086}$ & $.793_{.088}$ & $.595_{.056}$ & $.708_{.035}$ \\ 
    \hline[dotted] 
    
    CSMIL       & $.612_{.036}$ & $.619_{.048}$ & $.677_{.065}$ & $.806_{.067}$ & $.592_{.037}$ & $.716_{.035}$ \\ 
    \hline[dotted]
    
    \textbf{CrossFusion} & $.628_{.019}$ & $.684_{.043}$ & $.698_{.023}$ & $.810_{.067}$ & $.625_{.051}$ & $.745_{.030}$ \\ 
    
    \end{tblr}
    \label{tab:uni2}
\end{table*}

\rev{\subsection{Analysis of Inference Time}}

\rev{Inference Efficiency. We used the same 20x features in all methods. Our common feature extraction time was 30.61 seconds on average for 20x features in a Nvidia A4000 GPU.  For 10x and 5x features, the CLAM can extract features in parallel, which will take less than 10 seconds depending on the number of foreground objects in the image. So we compare the inference time here. CrossFusion demonstrates superior efficiency with a slide-level fusion time of just 0.145 seconds. By eliminating the heavy computational bottlenecks found in graph-based methods (e.g., Patch-GCN: ~40 seconds for graph construction) and hierarchical models (e.g., HIPT: ~55 seconds), our approach is significantly faster. This low-latency inference makes CrossFusion highly suitable for large-scale, real-time clinical deployment.
}

\section{Conclusion}

We introduced CrossFusion, a novel framework that fuses multi-scale patch embeddings from WSIs using cross-attention, transformer-based spatial encoding, and convolutional fusion. By integrating multi-scale features, CrossFusion captures key histopathological patterns linked to patient survival. Our experiments on diverse TCGA cancer datasets show that CrossFusion demonstrates significant improvements over the current state-of-the-art survival prediction methods, even under challenging conditions.

Our results underscore the value of domain-specific feature extraction in preserving crucial tissue details, such as cellular morphology and tissue architecture. The attention-based heatmaps further confirm the model's effectiveness and offer insights into its decision-making process.

In summary, CrossFusion bridges advanced deep learning with clinical needs, providing a robust and interpretable tool for cancer survival prediction. Future work will explore additional data modalities to guide the model to focus on important case-specific patterns and enhance interpretability, paving the way for more personalized cancer treatment and improved patient outcomes.

\clearpage

%
%
\bibliography{CrossFusion}

@inproceedings{ikezogwo2023quilt1m,
    title={Quilt-1M: One Million Image-Text Pairs for Histopathology},
      author={Ikezogwo, Wisdom Oluchi and Seyfioglu, Mehmet Saygin and Ghezloo, Fatemeh and Chan Geva, Dylan Stefan and Mohammed, Fatwir Sheikh and Anand, Pavan Kumar and Krishna, Ranjay and Shapiro, Linda},
      booktitle={Advances in Neural Information Processing Systems (NeurIPS)},
      year={2023}
    }

@article{lu2024avisionlanguage,
  author    = {Lu, Ming Y and Chen, Bowen and Williamson, Drew F K and Chen, Richard J and Liang, Ivy and Ding, Tong and Jaume, Guillaume and Odintsov, Igor and Le, Long Phi and Gerber, Georg and others},
  title     = {A visual-language foundation model for computational pathology},
  journal   = {Nature Medicine},
  volume    = {30},
  pages     = {863--874},
  year      = {2024},
  publisher = {Nature Publishing Group}
}

@inproceedings{ITW:2018,
    title={Attention-based Deep Multiple Instance Learning},
      author={Ilse, Maximilian and Tomczak, Jakub M and Welling, Max},
      booktitle={International Conference on Machine Learning (ICML)},
      pages={2127--2136},
      year={2018},
      organization={PMLR}
    }

@incollection{chen2021whole,
  author    = {Chen, Richard J and Lu, Ming Y and Shaban, Muhammad and Chen, Chengkuan and Chen, Tiffany Y and Williamson, Drew F K and Mahmood, Faisal},
  title     = {Whole Slide Images are 2D Point Clouds: Context-Aware Survival Prediction using Patch-based Graph Convolutional Networks},
  booktitle = {Medical Image Computing and Computer Assisted Intervention – MICCAI 2021},
  pages     = {339--349},
  publisher = {Springer International Publishing},
  year      = {2021},
  doi       = {10.1007/978-3-030-87237-3_33}
}

@inproceedings{10.1007/978-3-030-00934-2_20,
  author    = {Li, Ruoyu and Yao, Jiawen and Zhu, Xinliang and Li, Yeqing and Huang, Junzhou},
  title     = {Graph CNN for Survival Analysis on Whole Slide Pathological Images},
  editor    = {Frangi, Alejandro F and Schnabel, Julia A and Davatzikos, Christos and Alberola-L{\'o}pez, Carlos and Fichtinger, Gabor},
  booktitle = {Medical Image Computing and Computer Assisted Intervention – MICCAI 2018},
  pages     = {174--182},
  publisher = {Springer International Publishing},
  address   = {Cham},
  year      = {2018},
  isbn      = {978-3-030-00934-2}
}

@misc{yang2024scmilsparsecontextawaremultiple,
    title={SCMIL: Sparse Context-aware Multiple Instance Learning for Predicting Cancer Survival Probability Distribution in Whole Slide Images},
      author={Yang, Zekang and Liu, Hong and Wang, Xiangdong},
      booktitle={International Conference on Medical Image Computing and Computer-Assisted Intervention (MICCAI)},
      year={2024},
      organization={Springer}
    }

@inproceedings{li2021dualstreammultipleinstancelearning,
    title={Dual-stream Multiple Instance Learning Network for Whole Slide Image Classification with Self-supervised Contrastive Learning},
      author={Li, Bin and Li, Yin and Eliceiri, Kevin W},
      booktitle={Proceedings of the IEEE/CVF Conference on Computer Vision and Pattern Recognition (CVPR)},
      pages={14318--14328},
      year={2021}
}

@article{shao2021transmil,
  author    = {Shao, Zhuchen and Bian, Hao and Chen, Yang and Wang, Yifeng and Zhang, Jian and Ji, Xiangyang and others},
  title     = {Transmil: Transformer based correlated multiple instance learning for whole slide image classification},
  journal   = {Advances in Neural Information Processing Systems},
  volume    = {34},
  pages     = {2136--2147},
  year      = {2021}
}

@inproceedings{wu2024leveraging,
  author    = {Wu, Junxian and Ke, Xinyi and Jiang, Xiaoming and Wu, Huanwen and Kong, Youyong and Shao, Lizhi},
  title     = {Leveraging Tumor Heterogeneity: Heterogeneous Graph Representation Learning for Cancer Survival Prediction in Whole Slide Images},
  booktitle = {The Thirty-eighth Annual Conference on Neural Information Processing Systems},
  year      = {2024}
}

@article{lu2021data,
  author    = {Lu, Ming Y and Williamson, Drew F K and Chen, Tiffany Y and Chen, Richard J and Barbieri, Matteo and Mahmood, Faisal},
  title     = {Data-efficient and weakly supervised computational pathology on whole-slide images},
  journal   = {Nature Biomedical Engineering},
  volume    = {5},
  number    = {6},
  pages     = {555--570},
  year      = {2021},
  publisher = {Nature Publishing Group}
}

@inproceedings{He_2016_CVPR,
  author    = {He, Kaiming and Zhang, Xiangyu and Ren, Shaoqing and Sun, Jian},
  title     = {Deep Residual Learning for Image Recognition},
  booktitle = {Proceedings of the IEEE Conference on Computer Vision and Pattern Recognition (CVPR)},
  month     = {June},
  year      = {2016}
}

@article{chen2024uni,
  author    = {Chen, Richard J and Ding, Tong and Lu, Ming Y and Williamson, Drew F K and Jaume, Guillaume and Chen, Bowen and Zhang, Andrew and Shao, Daniel and Song, Andrew H and Shaban, Muhammad and others},
  title     = {Towards a General-Purpose Foundation Model for Computational Pathology},
  journal   = {Nature Medicine},
  year      = {2024},
  publisher = {Nature Publishing Group}
}

@article{xu2024gigapath,
  author    = {Xu, Hanwen and Usuyama, Naoto and Bagga, Jaspreet and Zhang, Sheng and Rao, Rajesh and Naumann, Tristan and Wong, Cliff and Gero, Zelalem and Gonz{\'a}lez, Javier and Gu, Yu and Xu, Yanbo and Wei, Mu and Wang, Wenhui and Ma, Shuming and Wei, Furu and Yang, Jianwei and Li, Chunyuan and Gao, Jianfeng and Rosemon, Jaylen and Bower, Tucker and Lee, Soohee and Weerasinghe, Roshanthi},
  title     = {A whole-slide foundation model for digital pathology from real-world data},
  journal   = {Nature},
  year      = {2024},
  publisher = {Nature Publishing Group UK, London}
}

@article{kumar2020whole,
  title={Whole slide imaging (WSI) in pathology: current perspectives and future directions},
  author={Kumar, Neeta and Gupta, Ruchika and Gupta, Sanjay},
  journal={Journal of digital imaging},
  volume={33},
  number={4},
  pages={1034--1040},
  year={2020},
  publisher={Springer}
}

@article{kothari2013pathology,
  title={Pathology imaging informatics for quantitative analysis of whole-slide images},
  author={Kothari, Sonal and Phan, John H and Stokes, Todd H and Wang, May D},
  journal={Journal of the American Medical Informatics Association},
  volume={20},
  number={6},
  pages={1099--1108},
  year={2013},
  publisher={BMJ Publishing Group}
}

@article{ghaznavi2013digital,
  title={Digital imaging in pathology: whole-slide imaging and beyond},
  author={Ghaznavi, Farzad and Evans, Andrew and Madabhushi, Anant and Feldman, Michael},
  journal={Annual Review of Pathology: Mechanisms of Disease},
  volume={8},
  number={1},
  pages={331--359},
  year={2013},
  publisher={Annual Reviews}
}

@inproceedings{liu2025generating,
  author={Liu, Sitong and Liu, Kechun and Margolis, Samuel and Wu, Wenjun and Knezevich, Stevan R. and Elder, David E. and Eguchi, Megan M. and Elmore, Joann G. and Shapiro, Linda G.},
  title={Generating seamless virtual immunohistochemical whole slide images with content and color consistency},
  booktitle={Proceedings of the IEEE 22nd International Symposium on Biomedical Imaging (ISBI)},
  year={2025},
  pages={1--5}
}

@article{zhao2024less,
  title={LESS: Label-efficient multi-scale learning for cytological whole slide image screening},
  author={Zhao, Beidi and Deng, Wenlong and Li, Zi Han Henry and Zhou, Chen and Gao, Zuhua and Wang, Gang and Li, Xiaoxiao},
  journal={Medical Image Analysis},
  volume={94},
  pages={103109},
  year={2024},
  publisher={Elsevier}
}

@article{wu2021scale,
  title={Scale-aware transformers for diagnosing melanocytic lesions},
  author={Wu, Wenjun and Mehta, Sachin and Nofallah, Shima and Knezevich, Stevan and May, Caitlin J and Chang, Oliver H and Elmore, Joann G and Shapiro, Linda G},
  journal={IEEE Access},
  volume={9},
  pages={163526--163541},
  year={2021},
  publisher={IEEE}
}

@article{campanella2019clinical,
  title={Clinical-grade computational pathology using weakly supervised deep learning on whole slide images},
  author={Campanella, G. and Hanna, M. G. and Geneslaw, L. and Miraflor, A. and Silva, V. W. K. and Busam, K. J. and Brogi, E. and Reuter, V. E. and Fuchs, T. J. and Klimstra, D. S.},
  journal={Nature Medicine},
  volume={25},
  number={8},
  pages={1301--1309},
  year={2019},
  publisher={Nature Publishing Group}
}

@inproceedings{thandiackal2022differentiable,
  title={Differentiable zooming for multiple instance learning on whole-slide images},
  author={Thandiackal, Kevin and Chen, Boqi and Pati, Pushpak and Jaume, Guillaume and Williamson, Drew FK and Gabrani, Maria and Goksel, Orcun},
  booktitle={European Conference on Computer Vision},
  pages={699--715},
  year={2022},
  organization={Springer}
}

@inproceedings{marini2021multi,
  title={Multi-scale task multiple instance learning for the classification of digital pathology images with global annotations},
  author={Marini, Niccolo and Ot{\'a}lora, Sebastian and Ciompi, Francesco and Silvello, Gianmaria and Marchesin, Stefano and Vatrano, Simona and Buttafuoco, Genziana and Atzori, Manfredo and M{\"u}ller, Henning},
  booktitle={MICCAI Workshop on Computational Pathology},
  pages={170--181},
  year={2021},
  organization={PMLR}
}

@article{deng2024cross,
  title={Cross-scale multi-instance learning for pathological image diagnosis},
  author={Deng, Ruining and Cui, Can and Remedios, Lucas W and Bao, Shunxing and Womick, R Michael and Chiron, Sophie and Li, Jia and Roland, Joseph T and Lau, Ken S and Liu, Qi and others},
  journal={Medical image analysis},
  volume={94},
  pages={103124},
  year={2024},
  publisher={Elsevier}
}

@inproceedings{unipro,
  title={Distilled Prompt Learning for Incomplete Multimodal Survival Prediction},
  author={Xu, Yingxue and Zhou, Fengtao and Zhao, Chenyu and Wang, Yihui and Yang, Can and Chen, Hao},
  booktitle={Proceedings of the Computer Vision and Pattern Recognition Conference},
  pages={5102--5111},
  year={2025}
}

@inproceedings{mmp,
title={Multimodal Prototyping for Cancer Survival Prediction},
  author={Song, Andrew H and Chen, Richard J and Jaume, Guillaume and Vaidya, Anurag J and Baras, Alexander S and Mahmood, Faisal},
  booktitle={International Conference on Machine Learning (ICML)},
  year={2024}
}

@inproceedings{HIPT,
  title={Scaling vision transformers to gigapixel images via hierarchical self-supervised learning},
  author={Chen, Richard J and Chen, Chengkuan and Li, Yicong and Chen, Tiffany Y and Trister, Andrew D and Krishnan, Rahul G and Mahmood, Faisal},
  booktitle={Proceedings of the IEEE/CVF conference on computer vision and pattern recognition},
  pages={16144--16155},
  year={2022}
}

@article{tan2023multi,
  title={A multi-scale fusion and transformer based registration guided speckle noise reduction for OCT images},
  author={Tan, Zhiwei and Shi, Fei and Zhou, Yi and Wang, Jingcheng and Wang, Meng and Peng, Yuanyuan and Xu, Kai and Liu, Ming and Chen, Xinjian},
  journal={IEEE Transactions on Medical Imaging},
  volume={43},
  number={1},
  year={2023}
}
%



\clearpage

\appendix
\renewcommand{\thefigure}{S\arabic{figure}}
\setcounter{figure}{0}

\begin{figure}[h!]
    \centering
    \includegraphics[width=0.5\textwidth]{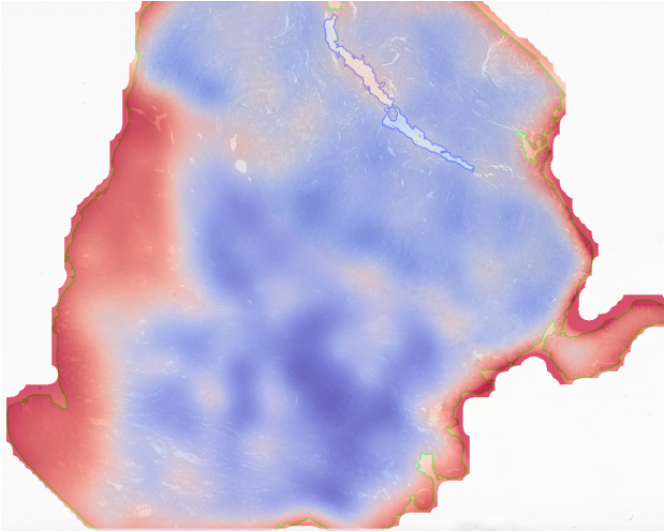}
    \caption{Failure Case - Low Risk.}
    \label{fig:fail1}
\end{figure}

\begin{figure}[h!]
    \centering
    \includegraphics[width=0.5\textwidth]{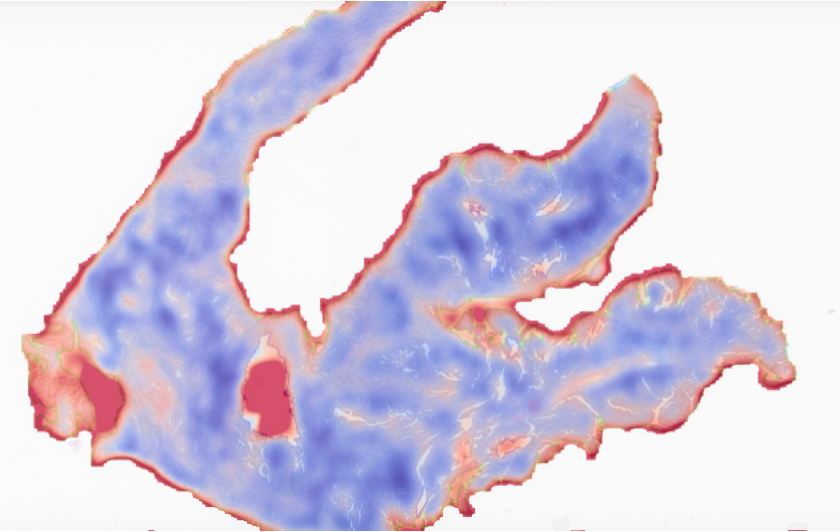}
    \caption{Failure Case - High Risk.}
    \label{fig:fail2}
\end{figure}

\reb{\section{Failure Case Analysis}}
\reb{To provide a comprehensive evaluation of CrossFusion, we analyzed cases where the model failed to correctly stratify patient risk. Figure \ref{fig:fail1} and  Figure \ref{fig:fail2} visualizes the attention heatmaps for two such representative failure cases.}

\reb{\textbf{Observation:} As observed in the heatmaps, the model demonstrates a distinct peripheral attention bias in these samples. The high-attention regions (highlighted in red) are predominantly concentrated along the boundaries of the tissue sections. Conversely, the core regions of the tissue—which typically harbor the dense tumor cells and critical morphological patterns required for accurate prognosis—are assigned low attention weights (indicated in blue).}

\reb{\textbf{Reasoning:} This behavior suggests that in these outliers, the model may have been distracted by slide preparation artifacts (e.g., tissue folding, edge compression, or marker residue) that frequently occur at the tissue periphery. Consequently, the model failed to capture the intrinsic tumor heterogeneity within the Region of Interest, leading to erroneous risk predictions. This finding highlights a potential direction for future improvement: incorporating boundary-aware regularization or more aggressive augmentation to suppress edge artifacts.}

\end{document}